\newcommand{\kms}{{\mbox{$\,\rm km~s^{-1}$}}}
\newcommand{\msun}{\,M_{\sun}\,}

\def\mathfont#1{\ifmmode{#1}\else{$#1$}\fi} 
\def\lae{\mathrel{<\kern-1.0em\lower0.9ex\hbox{$\sim$}}}  
\def\gae{\mathrel{>\kern-1.0em\lower0.9ex\hbox{$\sim$}}}

\def\qo{\ifmmode{q_o}\else{$q_0$}\fi}  
\def\ho{\ifmmode{H_o}\else{$H_0$}\fi}   
\def\hounit{\ifmmode{{\rm km\  s}^{-1}\ {\rm Mpc}^{-        
1}}\else{${\rm    km\ s}^{-1}\ {\rm Mpc}^{-1}$}\fi}  
\def\zf{\ifmmode{z_{GF}}\else{$z_{GF}$}\fi}  
\def\kms{\ifmmode{{\rm km\ s}^{-1}}\else{${\rm km\ s}^{-1}$}\fi}

\def\msun{\ifmmode{\ {\rm M}_\odot}\else{$ {\rm M}_\odot$}\fi}  
\def\msunyr{\ifmmode{\msun \ {\rm yr}^{-1}}\else{$\msun \ {\rm 
yr}^{-1}$}\fi}

\def\sfr{\ifmmode{\dot S}\else{$\dot S$}\fi}

\documentclass[
    ,final            % use final for the camera ready runs
  ]
  {aipproc}

\layoutstyle{6x9}

\begin{document}

\title{Calibrating AGN Feedback in Clusters}


\author{M. W. Wise}{
 address={ASTRON (Netherlands Institute for Radio Astronomy), \\ P.O. Box 2, 7990 AA Dwingeloo, The Netherlands},
 altaddress={Astronomical Institute Anton Pannekoek, University of Amsterdam, \\ Science Park 904, 1098 XH Amsterdam, The Netherlands}
}

\begin{abstract}
Whether caused by AGN jets, shocks, or mergers, the most definitive evidence for heating in cluster cores comes from X-ray spectroscopy. Unfortunately such spectra are essentially limited to studying the emission spectrum from the cluster as a whole. However since the same underlying emission measure distribution produces both the observed CCD and RGS spectra, X--ray imaging can still provide spatial information on the heating process. Using {\it Chandra} archival data for a sample of 9 clusters, we demonstrate how imaging data can be used to constrain departures from a canonical, isobaric cooling flow model as a function of position in a given cluster. The results of this analysis are also shown for the deep archival exposure of the Perseus cluster. Such ``heating maps'' can provide constraints on both the location and magnitude of the heating in the cores of clusters. When combined with detections and spectral index maps from low-frequency radio observations, these maps can be used to distinguish between different models for heating in these objects.
\end{abstract}

\maketitle

%%%%%%%%%%%%%%%%%%%%%%%%%%%%%%%%%%%%%%%%%%%%
%% MAINMATTER
%%%%%%%%%%%%%%%%%%%%%%%%%%%%%%%%%%%%%%%%%%%%

%
% Spectral evidence for heating in Clusters
%
\section{Introduction}
\label{sec:intro}

Since the launch of {\it Chandra} and {\it XMM-Newton}, it has become
evident that the X-ray emitting gas in the cores of galaxy clusters
is being heated and redistributed by powerful radio sources. Images
of cluster cores over the past few years show a wealth of structure,
much of which is produced by interactions between rapidly expanding
radio lobes and the hot intracluster (ICM) medium surrounding
them. In particular,  large cavities or bubbles are now
found routinely in the centers of clusters \citep{2007ARA&A..45..117M}.
Based on studies of cavity energetics, the global energy budget in these systems
seems more than sufficient to balance cooling in some objects 
\citep{2004ApJ...607..800B, 2006ApJ...652..216R}. 
The specifics of the feedback loop between AGN heating and cooling 
in cluster cores however is still not  understood. 
In particular, the details of where this energy deposition occurs 
and by what physical mechanism energy is transferred to the ICM 
remain open questions. 

In the canonical cooling flow model where gas is assumed to fully cool 
from ambient values down to very low temperatures, strong X--ray emission 
lines, such as Fe XVII at 15.02 \AA, would be expected. 
{\it XMM-Newton} RGS spectra of cluster cores however show a lack of strong 
X-ray emission lines from gas at temperatures below a few keV \citep{peterson:03a}.
Due to instrumental limitations, {\it XMM-Newton} RGS spectra are essentially 
limited to studying the emission spectrum from the cluster as a whole. 
Fortunately, X--ray imaging can sample the cluster temperature structure 
in cluster cores directly, albeit at reduced spectral resolution, 
since the same underlying emission measure distribution or
$dL/dT$ distribution produces both the observed CCD and RGS spectra. 
Using simple, parameterized functions for the $dL/dT$ distribution, 
we show here that one can map the departures from the standard isobaric 
cooling flow model as a function of position in a given cluster.  
These ``heating maps'' can provide constraints on both the location 
and magnitude of the heating in the cores of clusters.

%
% Mapping heating
%
\begin{figure}[t]
\hspace{-0.1in}
\includegraphics[width=6.0in,height=2.5in]{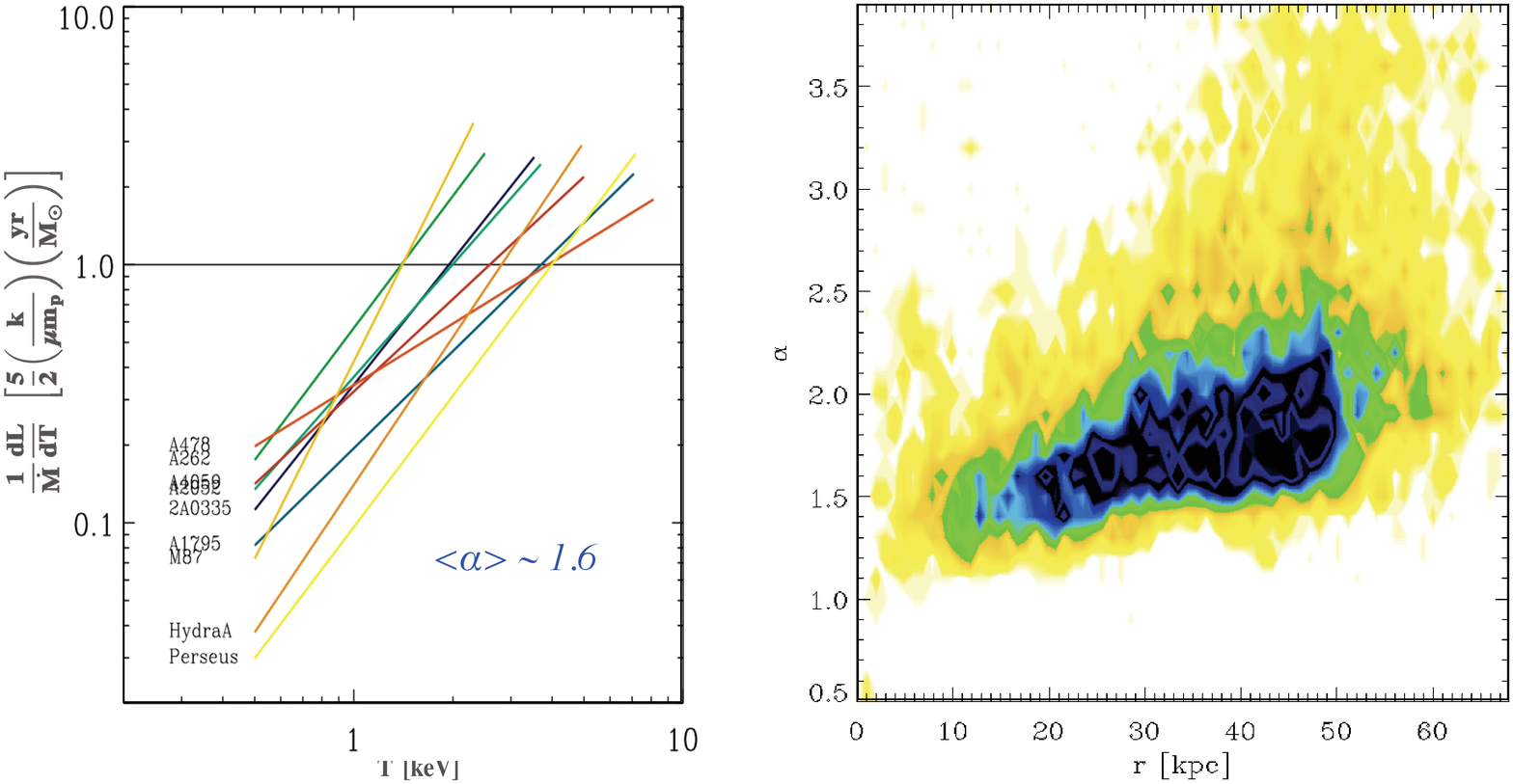}
\caption{\small
{\bf Left}: Integrated value of $\alpha$ determined within the central cooling region for a small sample of clusters with deep {\it Chandra} imaging data. This value represents a measure of the total heating rate in the core and can be compared to similar estimates based on {\it XMM-Newton} RGS spectra \citep{2009ApJ...emdist}.
{\bf Right}: Values of $\alpha$ as a function of radius in the Perseus cluster determined from archival {\it Chandra} imaging data. The increase in $\alpha$ with radius is indicative of an increase in heat deposition in the ICM \citep{2009ApJ...perseus}.}
\vspace{0.125in}
\label{fig:emdist}
\end{figure}

%
% Emission measure diagnostics
%
\section{Emission measure diagnostics}
\label{sec:emdist}

The lack of strong emission lines from gas at temperatures below 
$\sim T_o/3$, where $T_o$ is the ambient cluster temperature,
in {\it XMM-Newton} RGS spectra has become a key argument for new
cooling flow models which include such additional physics as AGN
heating, thermal conduction, and star formation.
In the standard cooling flow model, which assumes no heating, the
differential luminosity distribution, i.e. the luminosity radiated per
unit temperature interval, is proportional to the mass
deposition rate, $\dot {M}$:  
\begin{equation}
\frac{dL}{dT} = \frac52 \frac{k}{\mu m_p} \dot M ~~.
\label{eqn:isocool}
\end{equation}
Consequently in this scenario, one should observe equal amounts of
emission per unit temperature interval, including X--ray lines, from
gas down to very low temperatures. 
Based on a sample of 13 clusters, \cite{peterson:03a} find
a deficit of emission below about $\sim 1/3$ of the ambient
temperature relative to the predicted emission from 
equation~\ref{eqn:isocool}.
The observed distribution appears to be more consistent 
with an expression of the form,
\begin{equation}
\frac{dL}{dT} = \frac52 \frac{k}{\mu m_p} \dot M (\alpha +1) \left(
\frac{T}{T_0} \right)^{\alpha}
\label{eqn:nisocool}
\end{equation}
where $\alpha \sim 1$--$2$ instead of 0, as expected for the
standard cooling flow model.  In this prescription, $\alpha$ is 
directly proportional to the amount of heating in the cluster.
We have implemented the simple, parameterized $dL/dT$ distribution
model of equation~\ref{eqn:nisocool} as a source model within both the {\tt XSPEC}
and {\tt ISIS} spectral fitting packages \citep{arnaud:96a, 2000adass...9..591H}.
This technique is similar to that of \cite{kaastra:04a}, although they 
employ a different formalism for the $dL/dT$ distribution.

With this model, we have used archival {\it Chandra} imaging data to measure the 
underlying $dL/dT$ distribution for a small sample of clusters\footnote{This initial sample includes Perseus, M87, Hydra A, A478, A262, 2A0335+091, A2052, and A4052.}.
Several of these initial objects are common to the original sample of \cite{peterson:03a}.
Data were extracted and fit within the entire cooling radius for these objects so these results 
can be compared with the integrated RGS spectra directly.
The resulting $dL/dT$ distributions are shown plotted in the left panel of Figure~\ref{fig:emdist}.
The mean value of $\alpha$ for the sample is $<\alpha> \sim 1.6$.
These integrated results are completely consistent with the RGS results reported 
by \cite{peterson:03a}. We plan to expand this analysis to a larger sample 
of  {\it Chandra} and {\it XMM-Newton} imaging data.

\section{Mapping Heating in Cluster Cores}
\label{sec:maps}

\begin{figure*}[ht]
\hspace{-0.1in}
\includegraphics[width=5.0in]{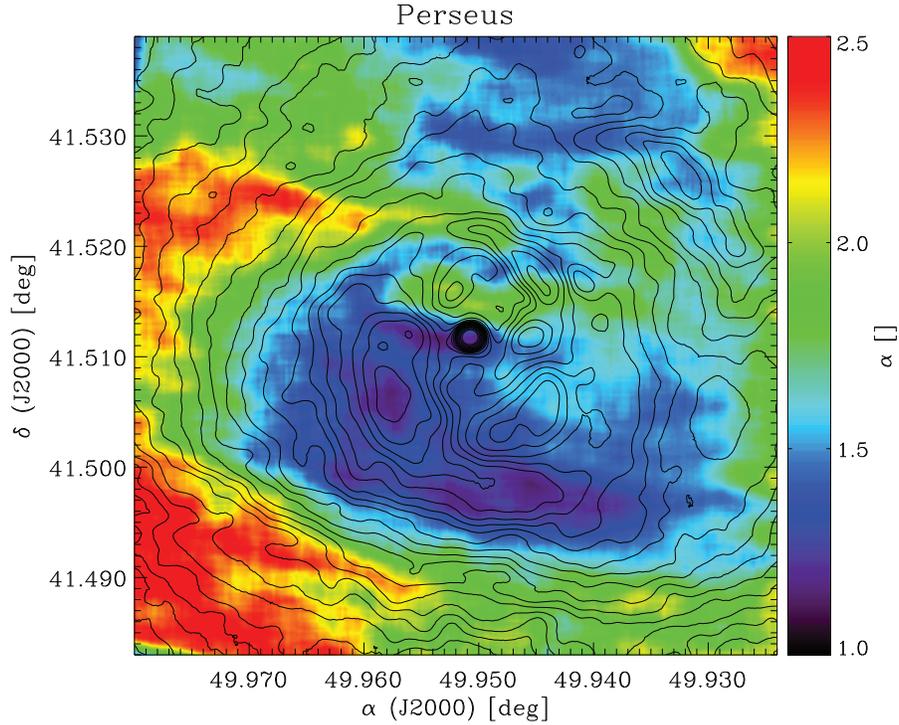}
\caption{\small A map of heating in the core of the Perseus cluster. The map shows the variation of the index $\alpha$, which measures the departure from simple, isobaric cooling, across the inner 50 kpc. The value of $\alpha$ at each point in the map is determined by fitting a power-law emission measure distribution using the 1 Msec of archival {\it Chandra} data on Perseus. Regions of rapidly changing $\alpha$ represent sites of heat deposition \citep{2009ApJ...perseus}.}
\vspace{0.125in}
\label{fig:heatmap}
\end{figure*}

Although fits to the integrated spectra make a nice consistency check, 
our ultimate goal is to actually map the departures from a
canonical, unheated cooling flow model (equivalent to determining the
$\alpha$ parameter) as a function of position in a given cluster. 
Figure~\ref{fig:heatmap} shows one such map constructed using the deep,
1 Msec  {\it Chandra} observation of the Perseus cluster.
It is immediately obvious that $\alpha > 0$ over the entire cooling
region. More strikingly, $\alpha$ rises rapidly immediately outside the
edges of the radio filled cavities. This observed spatial correlation
is strong circumstantial evidence that energy associated with the
radio cavities is being deposited into the ICM at these locations. 
Due to the episodic nature of AGN activity, the ambient, non-zero
value of $\alpha$ may represent the time-averaged result of several
outbursts, while the observed rise in $\alpha$ at the cavity
boundaries may be due to ongoing heating associated with the most
recent outburst. 

Again, although disconnected from the physics of the heating
mechanism, such ``heating maps'' can provide constraints on both the 
location and magnitude of the heating in the cores of clusters. 
As Figure~\ref{fig:heatmap} demonstrates, this technique provides
a potentially powerful diagnostic for determining the underlying
$dL/dT$ distribution in clusters and by extension the degree and
location of heating in the ICM. 
Using this technique, we intend to construct such heating maps 
for all clusters in the  {\it Chandra} and {\it XMM-Newton} archives 
with sufficient quality data. Ultimately, these maps can provide 
constraints on AGN feedback-based heating models as well as 
the energy transfer mechanism itself.
The complex radio bubble and X--ray cavity morphologies
observed in cluster cores make it clear that spatial information
is key to understanding AGN heating in the ICM.

%
% Mapping heating
%

%%%%%%%%%%%%%%%%%%%%%%%%%%%%%%%%%%%%%%%%%%%%%%%%
%%
%% Bibliography
%%
%%%%%%%%%%%%%%%%%%%%%%%%%%%%%%%%%%%%%%%%%%%%%%%%

\bibliographystyle{aipproc}   % if natbib is available
\bibliography{references}

\end{document}